\documentclass[a4paper,20pt]{article}
    \usepackage[T1]{fontenc}
    \usepackage{graphicx}
    \usepackage{amsmath}
    \usepackage{amsfonts}
    \renewcommand{\abstract}{}
\begin{document}
\makeatletter
\renewcommand{\@oddhead}{\textit{YSC'14 Proceedings of Contributed Papers} \hfil \textit{K. B\k{a}czek, B. Wszo\l ek}}
\renewcommand{\@evenhead}{\hfil \textit{K. B\k{a}czek, B. Wszo\l ek}}
\renewcommand{\@evenfoot}{\hfil \thepage \hfil}
\renewcommand{\@oddfoot}{\hfil \thepage \hfil}
\fontsize{11}{11} \selectfont

\title{ The Jagiellonians and the Stars }
\author{\textsl{K. B\k{a}czek$^{1}$, B. Wszo\l ek$^{2}$}}
\date{}
\maketitle
\begin{center} {\small{ $^{1}$ Jagiellonian University, Institute of Geography, ul. Go\l  \k{e}bia 8, Krak\'ow, Poland \\
$^{2}$ Jan D\l ugosz Academy, Institute of Physics, al. Armii
Krajowej 13/15, 42-200 Cz\k{e}stochowa , Poland \\ karrzw@poczta.onet.pl,
bogdan@ajd.czest.pl}}
\end{center}

\begin{abstract}
The largest centre for astronomical and astrological study in the
fifteenth century was the University of Cracow, which always was
under special care of Jagiellonians. The use of astronomy and
astrology at Jagiellonian courts in the fifteenth and sixteenth
centuries were very common. We try to convince the reader about
this, exposing very limited historical sources.
\end{abstract}

\section*{Introduction}

\indent \indent In the year 1364 King Casimir the Great received the
permission of the Pope to establish a university in the capital of
his kingdom, Cracow. The site of the university was, in all
probability, under the eye of the King in the royal castle on Wawel
Hill. The premature death of King Casimir in 1370 together with the
total lack of interest in the University by his successor, King
Ludvic the Hungarian, led to the collapse of the University.

\indent In 1384, Hedvig, the 11-year-old girl was called to Poland
by the knights and representatives of towns to ascend the Polish
throne. The magnates chose to give her hand to the pagan ruler of
the Grand Duchy of Lithuania, Jogaila, under the condition that
Lithuanians had to become Christian and part of the Polish Kingdom.
The Union was concluded at Krewo in 1385. One year later, Jogaila
was baptized in Cracow, assuming the name Ladislas, and the assembly
of Polish knights elected him King of Poland. This was the beginning
of the rules of Jagiellonian dynasty in Polish Kingdom.

\indent The University was refounded in the year 1400 through the
efforts of Queen Hedvig. Her pleas at the papal court in Avignon and
the bequeathing of her personal fortune to the School enabled it to
be re-established, on a new basis, as a full four-faculty medieval
university. Collegium Maius was built then, whereas circa 1405 a
chair of astronomy was founded by Stobner.  54 years later a chair
of astrology was created, which in the following years was famed
beyond the borders of the Polish Kingdom.

\indent The reinvigorated Cracow University immediately established
itself in the world of learning. Its first rector Stanis\l  aw of
Skarbimierz, the author of the famous work `De bello iusto', today
is regarded as one of the founders of international law. In the
second half of the fifteenth century the Cracow schools of
mathematics and astrology blossomed. Their most important
representatives were: Marcin Kr\'ol of \.Zurawica (1422- before
1460), Marcin Bylica of Olkusz (1433-1493) who became the chief
astrologer to King Maciej Korwin in Buda, Marcin Biem (circa
1470-1540) who devised the reform of the Julian calendar, Jan of G\l
og\'ow (1445-1507) who was the author of many mathematical and
astronomical tracts which were known to all of Europe, Wojciech of
Brudzewo (circa 1446-1495) who was the teacher of many students who
later went on to become leadind academics at other European
universities. During this period Miko\l aj Kopernik (Nicolaus
Copernicus) studied liberal arts in Cracow from 1491 to 1494. In his
later years he maintained that he was greatly indebted to the Cracow
University. The importance of this University can be gauged by the
fact that in the years 1433-1510 as many as 44\% of the students
came from abroad \cite{markowski}

\indent In the course of the 15th century representatives of Cracow
University radically changed their attitude toward astrology. While
distinguished doctors of canon law such as Stanis\l  aw of
Skarbimierz, Tomasz of Strzempin and theology professor Benedykt
Hesse distanced themselves from astrology and questioned the
reliability of astrological prognostications, at the end of the
century Jan of G\l  ogów, master of philosophy, regarded astrology
as a science and even obtained large profits from practising it,
although he realized that astrological practices conflicted with
Catholic teaching \cite{bull}.

\section*{Astrology in the Middle Ages}
\indent \indent It should be understood that medieval astrology was
not held universally to be an example of `diabolical' trumpery,
despite the disapproval it earned from certain fathers of the Church
such as St. Augustine. In the fifteenth and sixteenth century for
many people it was `proper science', which often meant the same as
astronomy and involved detailed observation of the skies, the
calculation of time and calendars (predicting in advance the date of
Easter, to give its most common and necessary use). It was also used
to establish celestial positions which were regarded by serious
scholars as being significant for their influence human life, for
birth, death, sickness, personal character and so forth. Cardinal
Pierre d'Ailly (1350-1420) attempted to set down a world chronology
from Creation to the Second Coming basing on astrological
calculations, and the Polish astrologer Jan of G\l og\'ow agreed
with him, that  astrology was not out of keeping with Christianity,
even though some bad astrologers might abuse it by mixing it with
magic \cite{malewich}. Therefore, the use of astrology at court was
not itself sinful (after \cite{rowell}).

\section*{Astrology and history}
\indent \indent As some historians argue, astrological writings are
very helpful when they want to do dating of many events in the past.
The famous annalist, Jan D\l  ugosz, was compiling his historical
chronicle of Polish Kingdom using old almanacs written by
astrologers. D\l  ugosz probably used astrological calendars on
which many historical events were recorded. There were made also
historical horoscopes to illustrate when certain cities or states
formed or expanded their borders \cite{rowell}.

\section*{Astrological predictions}
\indent \indent There are numerous arguments that astronomy and
astrology was not strange for Jagiellonians. They not only supported
the development of these sciences but also tried to relay on their
results and predictions. Astrology was known and practised at the
Jagiellonian court.

\indent Two horoscopes for the child of queen Hedwig are the oldest
preserved in Poland \cite{rosinska}. The first horoscope, incomplete
and with many corrections, refers to date of the conception of the
child, 16 IX 1398, at noon, for which ascendant falls in the
$6^{th}$ grade of Sagittarius. Horoscope was written just after this
date. The second horoscope refers to the date of birth (22 VI 1399,
at 11:28). Ascendant was in the $23^{rd}$ grade of Virgin. It was
written directly after childbirth, for it has no name of the
daughter. Astrologer predicted the birth of a son in 1399. When on
$22^{nd}$ June 1399 instead of the son, the daughter was born, the
Cracovian astrologer explained that he was wrong not the science.

\indent During the reign of Jogaila people started to pay careful
attention to celestial phenomena. Observations of Solar and Lunar
eclipses were carried out. During those days, the Battle of Grunwald
between the joined forces of the Kingdom of Poland and the Grand
Duchy of Lithuania against the Knights of the Teutonic Order took
place on July 15, 1410. From an anonymous chronicle we know that in
the night of $14^{th}$ and $15^{th}$ July the Moon had a crimson
colour and on its background a red sword was shown. At the face of
the Moon people  saw the battle between the Royal Army and Teutonic
Order and this battle ended with a great victory of the king while
the Teutonic Order were knocked off the Moon.

\indent In 1424-1427, an astronomer Henry Czech stayed at the court
of Jogilla. He was present during the childbirth of three sons of
Jogilla and predicted their future.

\indent On  March 2, 1456 a son of Casimir IV Jagiellon and queen
Elizabeth was born. The Royal couple had an astrologer, Peter
Gaszowiec (1430-1474), who prepared the horoscope for him.

\indent The period of Sigismund August (1548-1572) was a very good
time for astrologers. The King was enamoured in secret sciences.
Also his mother - queen Bona - tried to predict the future. Royal
astrologers of her son were Peter Proboszczowicz and Martin Fox.
After advice of Proboszczowicz, the King rescheduled a coronation
Catherine of Austrian two times. Astrologer also predicted the death
of the King. He mentioned the number 72 but did not say if this date
was the calendar year or an age of the King. In fact Sigismund died
in 1572 being 52. Martin Fox also predicted the death Sigismund but
in 1552. This prediction was nearly true. That year the King met
with Albrecht Hohenzollern. While entering a town the King was
welcome with cannon salvos. One of the cannonballs flew in the
direction of the King killing the person next to him. Proboszczowicz
and Fox had a big influence on the king.

\section*{Conclusions}
\indent \indent During the reigns of the Jagiellonian dynasty the
sciences referred to stars and the heaven were sufficiently
supported by the government. Astronomy and astrology were very
important in the XV-XVI centuries in the Polish Kingdom. In that
time Cracow was the main astronomical centre in Europe.  Astronomy
and astrology strongly influenced the public life and they were the
subject of great interest of the Church and government authorities.

\end{document}